\documentclass[floatfix,nofootinbib,prd,amsfonts]{revtex4-1}

\usepackage{graphicx}
\usepackage{epic}
\usepackage{eepic}
\usepackage{latexsym}

\newcommand{\eq}[1]{(\ref{#1})}
\newcommand{\be}{\begin{equation}}
\newcommand{\ee}{\end{equation}}
\newcommand{\bea}{\begin{eqnarray}}
\newcommand{\eea}{\end{eqnarray}}

\newcommand{\vs}[1]{\vspace{#1 mm}}
\newcommand{\hs}[1]{\hspace{#1 mm}}

\newcommand{\vacr}{|0,t_0\hs{-1}>}
\newcommand{\vacl}{<\hs{-1}0,t_0|}

\newcommand{\fc}{\phi_{cl}}
\newcommand{\hf}{\hat{\phi}}
\newcommand{\fb}{\phi_{b}}

\def\b{\beta}

\def\d{\delta}

\def\e{\epsilon}

\def\fr{\frac}

\def\l{\lambda}

\def\m{\mu}

\def\o{\omega}

\def\del{\partial}

\let\bm=\bibitem
\def\nn{\nonumber}

\begin{document}

\title{Stationary Phase Approximation and Instanton-like States for \\ Cosmological In-In Path Integrals}

\author{Ali Kaya}
\email[On sabbatical leave of absence from Bo\~{g}azi\c{c}i University. \\]{ali.kaya@boun.edu.tr}
\affiliation{Department of Physics, McGill University, Montreal, QC, H3A 2T8, Canada\\
and\\
Bo\~{g}azi\c{c}i University, Department of Physics, 34342, Bebek, \.{I}stanbul, Turkey\vs{15}}

\begin{abstract}

The path integral, which generates in-in correlation functions of a scalar field in a cosmological  spacetime, is shown to admit nontrivial classical solutions as stationary phases.  Although the solutions exist for Lorentzian signature, their contribution to the path integral is reminiscent  that of the instantons in Euclidean field theories. When the scalar potential has more than one locally stable vacua, the correlation functions receive contributions from all of them via these instanton-like configurations, which is similar to tunneling. We present  some explicit solutions for toy models and discuss possible implications of our results. 

\end{abstract}

\maketitle

\section{Introduction}

Understanding the physics of the early universe is both a challenge and an opportunity. With the recent observational advances in cosmology, one may hope to test new theories on energy scales that can never be reached by conventional accelerators and discover signs of various novel ideas. This also motivates one to have a comprehensive understanding of conventional approaches.     

Inflation is the most attractive paradigm in solving the problems of the standard big-bang model and its main predictions are in agreement with  observations so far. However, most of these predictions are based on semiclassical reasoning and a better understanding of inflationary theories in terms of well established physics is necessary for their robustness. In recent years,  following the work \cite{w1,w2}, there has been a growing interest in calculating quantum loop corrections to cosmological correlation functions during inflation. Earlier, various cosmological implications of (loop) quantum effects have been studied  (see e.g. \cite{e00,y1,e01,y2,e2,y3,e1,e0,y4,e3,e4,e5,e6}). One of the main reasons for the recent interest is the observation that interactions  give rise to primordial non-gaussianities \cite{mal}, which can potentially be  observed (see \cite{rev} for a review).  

As usual, the main technique in evaluating quantum corrections is the (in-in) perturbation theory. Perturbative loop calculations corresponding to  massless fields in inflationary spacetimes are plagued by infrared divergences and there are different views in the literature whether infrared effects are real or not (see e.g. \cite{iy1,ir1,iy2,iy3,ir2,ir3,iy4,iy5,ir4,ir5,iy6, iy7,iy8,ir6,ir7}, see also \cite{so1,so2,so3}  for earlier work). On the other hand, loop contributions may also be anomalously sensitive to the UV cutoff as discussed in \cite{uv}. There are some attempts to get  non-perturbative results such as the stochastic approach \cite{s1,s2,s3,s4,s5}. There are also some attempts to go beyond the one loop approximation \cite{bl1,bl2,bl3} or to work out the complete one-loop  effective action in a time dependent  background \cite{epy1,epy2,epy3,epy4,ep1,ep2,ep3}. 

In this paper, we consider the in-in path integral for the generating functional of cosmological correlations corresponding to an interacting scalar field in a Freedman-Robertson-Walker (FRW) spacetime. For a field theory defined in the flat space, instantons, if they are admitted,  give valuable non-perturbative information about the vacuum structure of the theory (or one may say that instantons exist if vacuum is nontrivial).  Our aim here is to see whether similar instanton-like classical solutions exist, which would correspond to nontrivial saddle points of the cosmological in-in path integrals. In searching for instanton solutions, one naturally makes a Wick rotation to the Euclidean signature \cite{col1}. However, it is not known how to Wick rotate a general cosmological spacetime. As an exception, if one considers the de Sitter space in conformal coordinates, which  is appropriate for Wick rotation due to the special form of the metric, one sees that Euclidean continuation does not give an "inverted" scalar potential. Therefore, efforts of constructing Euclidean solutions directly reminiscent  of the usual instantons  would fail.\footnote{Here, we only consider quantum fields in a fixed background. There exits  gravitational instantons  in the context of Euclidean quantum gravity in three dimensional de Sitter space \cite{malo}.}

On the other hand, a cosmological in-in path integral is very different in a few important ways from an in-out path integral defined in flat space. Namely;\\
(1) The boundary conditions are different compared to the in-out path integration. \\
(2) In applying saddle point approximation to an in-out path integral, one searches for classical configurations which have finite action to have a well defined expansion scheme. However, in-in path integral contains two action terms in the exponential with different signs and there is a possibility of cancelation even for configurations with "infinite" action.  \\
(3) The cosmic expansion changes the classical dynamics in a crucial way.\\
Thus, one may still search for classical configurations corresponding to stationary phases of an in-in path integral in the Lorentzian signature. As we will discuss, all the three properties listed above will be important in applying a rigorous  stationary phase approximation.  

Indeed, we will be able to show that when the scalar potential have more than one locally stable vacua, there are nontrivial classical configurations which correspond to stationary phases of the in-in path integral for the generating functional. Due to the existence of these solutions,  the correlation function receives contributions from all of the vacua. For instance, the vacuum expectation value of the scalar (1-point function) becomes the sum of the all of the minima of the potential even when the in-vacuum is specified around one of the vacua, which is like a tunneling effect.

As we will discuss, when the in-vacuum state is defined not at the asymptotic past infinity but at a finite time in the past, there appears some additional technical difficulties mainly related to the identification of the vacuum and the correct set of boundary conditions one must employ. We show that with some reasonable assumptions it is possible to construct instanton-like solutions for that case also. 

The plan of the paper is as follows. In the next section, we review the in-in path integral formalism for an interacting scalar field propagating in a cosmological spacetime. Most of the material presented in that section is well known. As partially new results, we give a slightly different derivation of the equivalence of the path integral and operator formalisms and we discuss Wick rotation for a scalar in de Sitter space (see also  \cite{an} for a rigorous discussion of Wick rotation). In section \ref{3}, we fix boundary  conditions and apply stationary phase approximation to determine the equations obeyed by saddle points. It turns out that the results alter if  the in-vacuum state is define at finite or infinite times. In that section, we give several examples and elaborate physically on why instanton-like solutions do not exist in some situations and why they are admitted in some others. In conclusions, we discuss possible implications of our results and indicate some open problems. 

\section{Review of the in-in path integral formalism}

In this section, we would like to review the path integral derivation of the in-in correlation functions. Our aim is to fix our notation and discuss Wick rotation to the Euclidean signature as a first attempt to search for instanton solutions in de Sitter space. The background metric is taken as 
\bea
ds^2&=&-dt^2+a(t)^2(dx^2+dy^2+dz^2)\nn\\
&=&a(\eta)^2(-d\eta^2+dx^2+dy^2+dz^2),\label{met}
\eea
where $t$ and $\eta$ are the proper and the conformal time coordinates, respectively. The scale factor of the de Sitter space is given by $a=\exp(H t)$ or $a=-\eta_0/\eta$, where $\eta_0=1/H$. Most of our considerations below will be valid for an arbitrary scale factor $a(t)$ and we will freely switch between these two coordinate systems. The action of a minimally coupled real scalar field propagating in this background is given by
\be\label{a}
S[\phi,J]=-\fr12\int d^4x\sqrt{-g}\left[\nabla_\m\phi\nabla^\m\phi+V(\phi)\right]-2J\phi,
\ee
where we introduce an external source $J$, which can be turned on or off. In this paper we only consider real scalar fields with this canonical action. 

\subsection{General theory} 

Our main object of interest is to calculate the vacuum expectation value 
\be\label{c}
\vacl Q(t)\vacr,
\ee
where $Q(t)$ is a polynomial of the field variables and $\vacr$ is the ground state of the system at time $t_0$. Preferably, one would let $t_0\to -\infty$, however in a realistic scenario $t_0$ can be finite as well (in that case there is a certain uncertainty even in the specification of the free vacuum \cite{un}). One should also work with the vacuum of the interacting theory and thus in perturbation theory corrections to the initially chosen free vacuum state must be taken into account.  

The most straightforward way to introduce a path integral for \eq{c} is to consider the generating functional 
\be\label{g}
Z[J^+,J^-]=\int D\phi  \vacl  \phi,t\hs{-1}>_{J^-}<\hs{-1}\phi,t \vacr_{J^+},
\ee
where $\int D\phi  |\phi,t\hs{-1}><\hs{-1}\phi,t|$ is the identity operator constructed from the field variables at time $t$ and the inner products in \eq{g} are evaluated in the presence of two independent external sources $J^+$ and $J^-$ coupled to field variabes.  Differentiating \eq{g} with respect to $J^+$ or $J^-$ at time $t$, and setting $J^+=J^-=0$  gives \eq{c}. It is easy to write the transition amplitudes in \eq{g} in terms of path integrals to obtain
\be\label{gb}
Z[J^+,J^-]=\int D\phi  \int \prod_{t_0}^t  {\cal D} \phi^+ {\cal D}\phi^-e^{iS[\phi^+,J^+]-iS[\phi^-,J^-]} \Psi_0[\phi^+(t_0)]\Psi_0^*[\phi^-(t_0)] ,
\ee
where  $\phi^+$ and $\phi^-$  integrals are over all field configurations starting from time $t_0$ and ending at time $t$ obeying 
\be\label{bc}
\phi^+(t,\vec{x})=\phi^-(t,\vec{x})=\phi(\vec{x}),
\ee
and $\Psi_0[\phi^\pm(t_0)]$ are the vacuum wave-functionals corresponding to the inner products $<\hs{-1}\phi^\pm(t_0)\vacr$.  By introducing a Dirac-delta functional it is also possible to rewrite the path integral as
\be\label{gd}
Z[J^+,J^-]= \int  \prod_{t_0}^t {\cal D} \phi^+ {\cal D}\phi^-e^{iS[\phi^+,J^+]-iS[\phi^-,J^-]}  \Psi_0[\phi^+(t_0)]\Psi_0^*[\phi^-(t_0)]\delta[\phi^+(t)-\phi^-(t)] , 
\ee
where now there is no restriction imposed, and \eq{gd}  reduces to \eq{gb} after integrating over $\phi^+(t)$ or $\phi^-(t)$. 

\subsection{Perturbative expansion}

Before discussing the Wick-rotation in  Sitter space, it is instructive to see how perturbation theory works in the path integral formalism.\footnote{This issue is discussed in the appendix of \cite{w1}. Our treatment of the delta functional in \eq{gd} is different than  \cite{w1} but  its effect turns out to be the same, namely to force the appropriate boundary conditions on the propagators consistent with the operator formalism.} For that the free scalar action can be written as
\be
S_0[\phi,J]=\fr12\int d^4x \left[\phi L \phi\right]+2J\phi,
\ee
where $L$ is a (second order) differential operator. It is known that  the basic role of  the vacuum wave-functionals in a (free) path integral is to impose the proper $i\e$ prescription necessary to define a unique inverse of  $L$ (see section 9.2 of \cite{wb}). Therefore, using an integral representation of the delta functional in \eq{gd}, the free generating functional can be written as
\bea
&&Z_0[J^+,J^-]= \int D\l(\vec{x}) \int \prod_{t_0}^t {\cal D} \phi^+ {\cal D}\phi^-e^{iS_0[\phi^+,J^+]-iS_0[\phi^-,J^-]} \exp\left(i\int d^3x \l(\vec{x})\left[\phi^+(\vec{x},t)-\phi^-(\vec{x},t)\right]\right) \nn\\
&&= \int D\l(\vec{x}) \int \prod_{t_0}^t {\cal D} \Phi \exp\left(\fr{i}{2}\int_{t_0}^tdt'd^3x \left[ \Phi^T {\bf L}\Phi+\Phi^T{\cal J}\right]\right),\label{l}
\eea
where we define\footnote{Note that it is possible to shift  the upper limit of the time integral in \eq{l} by an arbitrary positive number since in-in formalism guarantees that the contributions for times larger than $t$ cancel. This justifies the delta functions in \eq{j}. We would like to thank the anonymous referee for pointing out this to us.}
\be\label{j}
{\bf L} =\left[\begin{array}{cc}L&0\\0&-L\end{array}\right],\hs{5}\Phi=\left[\begin{array}{c}\phi^+ \\\phi^- \end{array}\right],\hs{5} {\cal J}=\left[\begin{array}{c}J^+(\vec{x},t') + \l(\vec{x})\delta(t'-t) \\ -J^-(\vec{x},t')-  \l(\vec{x})\delta(t'-t) \end{array}\right].
\ee
Performing the Gaussian integral over $\Phi$ one  finds
\bea\label{z}
Z_0[J^+,J^-]= \int D\l(\vec{x}) \exp\left(-\fr{i}{2}\int_{t_0}^t dt' d^3x' \int_{t_0}^t dt'' d^3x''{\cal J}^T \Delta {\cal J}\right), 
\eea
where $\Delta$ is the inverse of ${\bf L}$ (with the $i\e$ prescription implied by the vacuum wave-functionals\footnote{Although in the free theory this procedure is relatively easy to implement, applying $i\e$ prescription for the interacting case using the in-in perturbation theory is highly non-trivial. See \cite{ie1,ie2} for clarification   of this point.}). One can write $\Delta$ as 
\be
\Delta=\left[\begin{array}{cc}\Delta^{++}&\Delta^{+-}\\ \Delta^{-+}&\Delta^{--}\end{array}\right],
\ee
where $L\Delta^{++}=1$, $L\Delta^{--}=-1$, $L\Delta^{+-}=0$ and $L\Delta^{-+}=0$. Since $L$ is a symmetric operator Green functions obey
\bea
&&\Delta^{+-}(\vec{x}',t';\vec{x}'',t'')=\Delta^{-+}(\vec{x}'',t'';\vec{x}',t'),\nn\\
&&\Delta^{++}(\vec{x}',t';\vec{x}'',t'')=\Delta^{++}(\vec{x}'',t'';\vec{x}',t'),\\
&&\Delta^{--}(\vec{x}',t';\vec{x}'',t'')=\Delta^{--}(\vec{x}'',t'';\vec{x}',t').\nn
\eea
Inside the exponential of the functional integral \eq{z}, there are quadratic and linear $\l$ terms. After performing the integrals over $t'$ and $t''$, the quadratic terms become
\bea
\exp\left(-\fr{i}{2} \int d^3x' d^3x''\l(x')\left[\Delta^{++}(t)+\Delta^{--}(t)-\Delta^{+-}(t)-\Delta^{-+}(t)\right]\l(x'')\right).
\eea
The Green function inside the square bracket is annihilated by $L$ and thus it is not invertible. Therefore, the only way to make  $\l$ path integral to be well defined is to impose these terms to cancel each other. In that case $\l$ integral becomes 
\bea
\int D\l \exp\left(-i \int d^3x' d^3x''\l(x')\left[[\Delta^{++}(t)-\Delta^{-+}(t)]J^+ +[\Delta^{--}(t)-\Delta^{+-}(t)]J^-\right]\right),
\eea
which is non-zero for arbitrary $J^+$ and $J^-$ provided
\bea
\Delta^{++}(t)=\Delta^{-+}(t),\nn\\
\Delta^{--}(t)=\Delta^{+-}(t).
\eea
These are the boundary conditions, which must be imposed on the homogenous solutions $\Delta^{+-}$ and $\Delta^{-+}$ to get a well defined path integral.  When these are imposed  $\l$ integral decouples and the generating functional becomes
\be
Z_0[J^+,J^-]=\exp\left(-\fr{i}{2}\int_{t_0}^t dt' d^3x' \int_{t_0}^t dt'' d^3x'' {\bf J}^T\Delta {\bf J}\right),
\ee
where 
\be
{\bf J}=\left[\begin{array}{c}J^+  \\ -J^- \end{array}\right]. 
\ee
Differentiating this final expression with respect to $J^+$ and $J^-$ one may verify from the definition of $Z$  that 
\bea
&&\Delta^{++}(\vec{x}',t';\vec{x}'',t'')=-i\vacl T\phi(\vec{x}',t')\phi(\vec{x}'',t'')\vacr, \nn\\
&&\Delta^{--}(\vec{x}',t';\vec{x}'',t'')=-i\vacl \overline{T}\phi(\vec{x}',t')\phi(\vec{x}'',t'')\vacr, \label{prop}\\
&&\Delta^{-+}(\vec{x}',t';\vec{x}'',t'')=-i\vacl \phi(\vec{x}',t')\phi(\vec{x}'',t'')\vacr,\nn
\eea
where $T$ and $\overline{T}$ deontes time and anti-time orderings, respectively. These are exactly the propagators obtained in the operator formalism. As usual, for an interacting theory with a polynomial potential $V(\phi)$, the generating functional can be expressed as
\be
Z[J^+,J^-]=\exp\left[\int -iV(-i\d/\d J^+)/2\right]\exp\left[i\int V(i\d/\d J^-)/2\right]Z_0[J^+,J^-],
\ee
which can be evaluated order by order using perturbation theory. 

\subsection{de Sitter space and Wick rotation}

\begin{figure}
\centerline{\includegraphics[width=6cm]{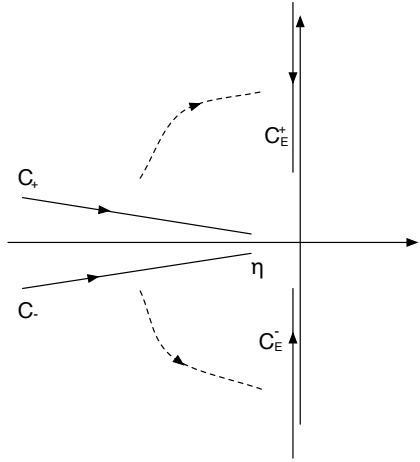}}
\caption{The integration contours $C_+$ and $C_-$ in the complex $\eta$-plane. The Wick rotated contours are also shown as $C_E^\pm$.} 
\label{fig1}
\end{figure}

At this moment, it is useful to discuss the specification of the vacuum state $\vacr$,  which is closely related to  Wick rotation. For that discussion, we specifically consider a real scalar field in the Poincare patch of the de Sitter space and use conformal coordinates $(\eta,\vec{x})$, where $\eta<0$. The easiest way to specify the interacting vacuum at $\eta_0=-\infty$ is to employ a projection by giving a small imaginary part to the time parameter $\eta$. In general, to project out an arbitrary ket-vector onto the interacting vacuum state defined at $\eta_0=-\infty$, one may introduce the operator $\exp(-\e H\Delta\eta)$ with $\Delta\eta\to\infty$ and $\e>0$, where $H$ is the exact Hamiltonian.\footnote{In the following discussion we take $H$ to be  time independent. With slight modifications, the arguments must be generalized to time-dependent situations, at least  in the adiabatic limit.} Since the unitary time evolution operator is given by $U=\exp(-iH\eta)$, such a projection can be naturally incorporated  by assuming that time parameter has a small negative imaginary piece. Therefore for fields evolving forward in time, which correspond to the $+$ branch in the in-in formalism, the time must be complexified as
\be\label{c1}
\eta_+=\eta_r(1-i\e),
\ee
where $\eta_r$ is real and $\e>0$. Similarly, to project out an arbitrary bra-vector onto the interacting vacuum, time must be  complexified as
\be\label{c2}
\eta_-=\eta_r(1+i\e). 
\ee
After these deformations of the integration contours, there is no need to keep the vacuum wave-functionals in \eq{gb} and one writes
\bea
Z[J^+,J^-]&=&\int D\phi  \int\prod_{C_+}{\cal D} \phi^+\int\prod_{C_-} {\cal D}\phi^-e^{iS[\phi^+,J^+]-iS[\phi^-,J^-]},  \nn
\eea
where $C_+$ and $C_-$  are defined by \eq{c1} and \eq{c2} in the complex $\eta$-plane (see  Fig. \ref{fig1}).   

Let us determine how introducing this  complex tilt affects the propagators of a massless field in de Sitter space. It is well known that the scalar field operator can be expanded in terms of the mode functions as
\be
\phi(\vec{x},\eta)=-H\eta\int \fr{d^3k}{(2\pi)^{3/2}}\fr{1}{\sqrt{2k}}\left[e^{i\vec{k}.\vec{x}-ik\eta}
\left[1-\fr{i}{k\eta}\right]a_k+e^{-i\vec{k}.\vec{x}+ik\eta}
\left[1+\fr{i}{k\eta}\right]a_k^\dagger\right].
\ee
In evaluating the field operator $\phi(\vec{x},\eta)$ on the contours $C_+$ and $C_-$, one should replace $\eta$ by $\eta_+$ and $\eta_-$, respectively, which gives
\be
\Delta^{-+}(\eta_-';\eta_+'')=\int d^3k\,e^{-ik\eta_-' +ik\eta_+''}...=\int d^3k\,e^{k\e(\eta_r' +\eta_r'')}...=\int d^3k\,e^{-k\e}... 
\ee
and 
\bea
\Delta^{++}(\eta_+';\eta_+'')&=&\theta(\eta'-\eta'')\int d^3k\,e^{-ik\eta_+'' +ik\eta_+'}...+\theta(\eta''-\eta') \int d^3k\,e^{-ik\eta_+' +ik\eta_+''}...\nn\\
&=&\theta(\eta'-\eta'')\int d^3k\,e^{k\e(\eta_+' -\eta_+'')}...+\theta(\eta''-\eta') \int d^3k\,e^{k\e(\eta_+'' -\eta_+')}...=\int d^3k\,e^{-k\e}... \,\, ,
\eea
where the signs are fixed by paying attention to the orderings along the respective contours and any positive number multiplying $\e$ is absorbed in $\e$. The crucial point is that properly complexifying the time coordinates  gives the extra $\exp(-k\e)$ factors in the momentum integrals, which are necessary for the UV convergence of the Green functions (see e.g. \cite{e2}). A similar calculation shows that the same damping factors appear for $\Delta^{+-}$ and $\Delta^{--}$, which become well defined in the UV.

The above discussion clearly indicates how one should perform Wick rotation to the Euclidean signature.\footnote{In  flat space, Wick rotation of the in-in path integrals has been discussed in \cite{eq}.} Since the propagators are well defined along the contours $C_+$ and $C_-$, and they are analytic functions of their complex variables provided $\e>0$, one can do the following replacements without changing the propagators:\footnote{In other words, by Wick rotation the oscillating functions damped by a small convergence factor are replaced by the exponentially decaying functions of the Euclidean time.}
\bea
\eta_+\to -i\eta_r,  \nn\\
\eta_-\to +i\eta_r. \nn
\eea
After this rotation, the Lorentzian actions for $+$ and $-$ branches 
\be
S_\pm[\phi^\pm]=\fr12\int_{C_\pm} d\eta d^3x\left[(\del_\eta\phi^\pm)^2-(\vec{\del}\phi^\pm)^2-\fr{V(\phi^\pm)}{\eta^2}\right]\fr{1}{\eta^2},
\ee
will be transformed into
\be
S_E^\pm[\phi^\pm]=\fr{i}{2}\int d\eta_r d^3x\left[(\del_{\eta_r}\phi^\pm)^2+(\vec{\del}\phi^\pm)^2-\fr{V(\phi^\pm)}{\eta_r^2}\right]\fr{1}{\eta_r^2}. 
\ee
Note that the relative signs of the kinetic and the potential energy terms are the same for the Lorentzian and the Euclidean actions since both signs are changed. Of course, this could be anticipated from the beginning since the sign change for the kinetic term arises due to time derivatives and the sign of the potential is changed due to the metric fuctions. However, it was necessary to work out the intermediate steps as we did above to make sure that analytical continuation can be performed without a problem.

As a result, we see that Wick rotation to the Euclidean signature in the Poincare patch of the de Sitter space does not give an inverted potential and therefore it is not possible to construct standard instanton solutions for the scalar fields, which are supposed to extrapolate in between different vacua as in the case of flat spacetime.\footnote{Note, however, that the relative signs of the kinetic and the spatial gradient terms are changed.} Nevertheless, we will see in the next section that it is possible to construct classical solutions in the Lorentzian  signature, which are reminiscent of  instantons. In the de Sitter space, if the scalar field is assumed to depend only on time, as we will suppose in the next section, then the Lorentzian and the Euclidean field equations will be the same due to the special property of the Wick rotation discussed above. Namely any Lorentzian solution would also satisfy Euclidean field equations and vice versa. Therefore, in de Sitter space these solutions also correspond to the saddle points of Euclidean path integrals similar to  instantons. 

It is useful to remind that our discussion in this subsection is carried out in the Poincare patch of the de Sitter space, which is relevant for cosmology. The full de Sitter space has different properties and viewing it as the hyperboloid in the Minkowski space of one higher dimension, the Wick rotation to the Euclidean signature gives a sphere. 

\section{Stationary phases and Instanton-like states}\label{3}

Consider the following path integral 
\be\label{i}
I=\int D\phi  \int \prod_{t_0}^t  {\cal D} \phi^+ {\cal D}\phi^-\exp\left(iS[\phi^+]-iS[\phi^-]\right) \Psi_0[\phi^+(t_0)]\Psi_0^*[\phi^-(t_0)] \,,
\ee
which appears in the generating functional \eq{gb}. Our aim is to see if stationary phase approximation can be used to evaluate \eq{i}. We take the scalar action \eq{a} defined in a general  FRW spacetime with the metric \eq{met}. There are three integration variables in \eq{i} and the phase must be stationary with respect to each of them. Moreover, although their existence only affects the integral at $t_0$, the presence of the vacuum wave-functionals must also be taken into account. 

\subsection{Boundary conditions}

Surface terms may arise in our discussion for two different reasons: either as a result of integrating by parts the field variables or from the variations of the action. The vanishing of the surface terms for the first case is important to have a well defined path integral. For example, in the free theory these integration by parts are necessary to obtain the second order differential operator in the action. Boundary conditions making these surface terms to vanish must be imposed on all paths contributing to the in-in path integral, defining the function space in which the functional integral is carried on. The surface terms in the second case will be important in searching for stationary phases of the path integral as we will discuss in the next subsection.  

Although it might be important to determine the precise boundary conditions in a more detailed study, here we simply assume that necessary conditions are imposed at {\it spatial} infinity and concentrate on the surface terms arising in the time direction. These surface terms can only arise from the kinetic term in \eq{a}. Here, one has the option of choosing two different alternatives. If one insists on freely integrating by parts the $+$ and  $-$ branches, the following conditions must be imposed on the fields separately
\be
\textrm{Strong conditions:}\hs{5} \fr{\del \phi^\pm}{\del t'}(t)=0,\hs{8}a^3(t_0)\phi^\pm(t_0) \fr{\del \phi^\pm}{\del t'}(t_0)=0.\label{s}
\ee
On the other hand, in many cases it would be enough to require the absence of surface terms for simultaneous integration by parts of the $+$ and $-$ fields, which implies 
\be\label{w}
\textrm{Weak conditions:}\hs{5}\fr{\del \phi^+}{\del t'}(t)-\fr{\del \phi^-}{\del t'}(t)=0,\hs{8}a^3(t_0)\left[\phi^+(t_0) \fr{\del \phi^+}{\del t'}(t_0)-\phi^-(t_0) \fr{\del \phi^-}{\del t'}(t_0)\right]=0.
\ee
As noted above, these boundary conditions determine  the function space on which the path integration is carried out. Our main results will not depend on the choice of the strong or the weak boundary conditions. However, it would be important  to find out the correct conditions for some applications. 

To continue, one must consider the infinite and the finite $t_0$ cases separately. For $t_0$ finite, we assume that $a(t_0)$ is well defined, i.e. we exclude the situations involving a big-bang singularity. For the other case $t_0=-\infty$, we will assume, having of course inflation in mind, that $a(t_0)\to 0$ as $t_0\to-\infty$ (which we simply write as $a(-\infty)=0$).  Let us first discuss $t_0=-\infty$ case for which the boundary conditions can be determined unambiguously. We would like to remind the reader that  the letter $t$ is reserved to denote the present time and we use $t'$ as a dummy variable if necessary. 

\subsection{The case $t_0=-\infty$}

The path integral \eq{i} is over all paths extending from $t_0=-\infty$ to time $t$ ($+$ branch) and then back to $t_0=-\infty$ ($-$ branch). Therefore, a stationary phase of \eq{i} is a path, conveniently named as $\Phi_{cl}(t',\vec{x})$, which has independent  $+$ and $-$ branches,  denoted by  $\fc^\pm(t',\vec{x})$, respectively (the branches are connected at time $t$). The variation around such a path can be parametrized by $\d\phi^+$ and $\d\phi^-$ obeying
 \be\label{vb}
 \d\phi^+(t,\vec{x})=\d\phi^-(t,\vec{x})=\d\phi(\vec{x}),
\ee
where $\d\phi(\vec{x})$ corresponds to the variation of the path at the "boundary" time $t$. The path $\Phi_{cl}$ is stationary if 
\be
\fr{\d}{\d\phi^\pm}\left(S[\phi^+]-S[\phi^-]\right)_{\Phi_{cl}}=0.
\ee
From this variation, the following surface term arises (as usual surface terms along spatial directions are assumed to vanish by suitable boundary conditions)
\bea\label{ara}
\lim_{t_0\to-\infty}\left[\d\phi^+a^3\fr{\del\fc^+}{\del t'}-\d\phi^-a^3\fr{\del\fc^-}{\del t'}\right]^t_{t_0}&=& \d\phi \left[\fr{\del\fc^+}{\del t'}(t)-\fr{\del\fc^-}{\del t'}(t)\right]a(t)^3\\
&-&\d\phi^+(t_0)\fr{\del \fc^+(t_0)}{\del t'}a(t_0)^3+\d\phi^-(t_0)\fr{\del \fc^-(t_0)}{\del t'}a(t_0)^3=0\nn,
\eea
where the condition \eq{vb} is used. Since $\d\phi$ is independent,  \eq{ara} implies $\del\fc^+(t,\vec{x})/\del t=\del\fc^-(t,\vec{x})/\del t$. Together with the boundary condition $\fc^+(t,\vec{x})=\fc^-(t,\vec{x})$ (recall that $\fc^+$ and $\fc^-$ denote two different branches of the same path $\Phi_{cl}$), one sees that the path corresponding to the stationary phase must obey
\be\label{ara2}
\fc^+=\fc^-\equiv\fc
\ee
for all times in the region$(-\infty,t)$, because they have the same initial value data.  The remaining terms in \eq{ara} vanish since we consider FRW spacetimes with $a(-\infty)=0$ and thus \eq{ara} is satisified. Thus, requiring that the phase to be stationary with respect to the variations of the boundary variable in \eq{i} implies the equality of the $+$ and $-$ paths. On the other hand, independent variations $\d\phi^\pm$ implies the same condition for $\fc$: 
\be\label{eq}
\fr{\d S[\fc]}{\d\phi}=0. 
\ee
i.e. $\fc$ must obey the classical equations of motion. 

One may see that $\Phi_{cl}$ satisfies all the weak boundary conditions \eq{w} and only the two of the strong boundary conditions \eq{s}. Therefore, if one insists on imposing the strong boundary conditions then $\del\fc(t,\vec{x})/\del t=0$ must also be satisfied. 

Till now we have not yet worked out the vacuum wave-functionals in \eq{i} which directly affects the integrations over $\phi^\pm(-\infty)$.  The vacuum wave-functional of the free theory is a Gaussian centered around $\phi=0$. The exact form of the vacuum wave-functional in an interacting theory is not known but in perturbation theory there must arise corrections to the Gaussian functional. In any case, if the theory is expanded around $\phi=0$, the vacuum wave-functionals are expected to be oscillating wave functionals around that point. Consequently, the stationary phase  approximation applied to $\phi^\pm(-\infty)$ integrations  implies
\be\label{sbc}
\fc(-\infty,\vec{x})= 0.
\ee
If one perturbatively  expands around a different vacuum, say $\phi=\phi_0$ corresponding to a minimum of the potential, then \eq{sbc} must be replaced  by $\fc(-\infty,\vec{x})=\phi_0$.

We thus conclude that any classical configuration obeying the equations of motion \eq{eq} and the asymptotic boundary condition \eq{sbc}  gives rise to a stationary phase of the integral \eq{i} (with an additional constraint $\del\fc(t,\vec{x})/\del t=0$ if strong boundary conditions are imposed).  Here, there only appears a single boundary condition \eq{sbc} and the situation is obviously different than the boundary conditions arising in a typical in-out path integral having two asymptotic regions. This is the point (1) mentioned in the introduction and it will be crucial in constructing classical solutions as stationary phases. 

Let us now expand the generating functional \eq{gb} around  a solution $\fc$. We define new integration variables as 
\bea
&&\phi^+=\fc+\hf^+,\nn\\
&&\phi^-=\fc+\hf^-,\\
&&\phi=\fb +\hf,\nn
\eea
where $\fb$ is the boundary value of $\fc$
\be
\fb(\vec{x})\equiv\fc(t,\vec{x}). 
\ee
In these new variables \eq{gb} becomes
\bea
&&Z[J^+,J^-]_{inst}=\int D\hf  \int \prod_{-\infty}^t  {\cal D} \hf^+ {\cal D}\hf^-e^{iS[\fc+\hf+,J^+]-iS[\fc+\hf^-,J^-]} 
\Psi_0[\phi^+(-\infty)]\Psi_0^*[\phi^-(-\infty)]\nn\\
&&=\exp\left[i\int \fc (J^+-J^-)\right]\int D\hf  \int \prod_{-\infty}^t  {\cal D} \hf^+ {\cal D}\hf^-e^{i\hat{S}[\fc;\hf^+,J^+]-i\hat{S}[\fc;\hf^-,J^-]}\hat{\Psi}_0[\hf^+(-\infty)]\hat{\Psi}_0^*[\hf^-(-\infty)],\label{ex}
\eea
where $\hat{\Psi}_0$ denotes the new vacuum wave-functionals and the hatted integration variables must obey 
\be\label{bcn}
\hf^+(t,\vec{x})=\hf^-(t,\vec{x})=\hf(\vec{x}).
\ee
The new action $\hat{S}$ contains quadratic and higher order powers of the field variable, which can be written explicitly as
\be
\hat{S}[\hf,J]=-\fr12\int d^4x\sqrt{-g}\left[\nabla_\m\hf\nabla^\m\hf+\hat{V}(\hf)-2J\phi\right],
\ee
where the new potential $\hat{V}$ is given by 
\be
\hat{V}(\hf)=V(\hf+\fc)-V(\fc)-V'(\fc)\hf.
\ee
The terms linear in $\hf^\pm$ cancel out after an integration by parts since $\fc$ obeys equations of motion and surface terms vanish owing to the boundary conditions \eq{bcn}. 

Differentiating with respect to $J^+$ and setting all external sources to zero one finds from \eq{ex} that 
\be\label{nz}
<\phi>=\fc,
\ee
where we assume that $\hf=0$ is a minimum of $\hat{V}(\hf)$ and the path integral in \eq{ex} is perturbatively evaluated around this vacuum implying $<\hf>=0$.  A non-zero vacuum expectation value such as \eq{nz} cannot be generated in perturbation theory. 

It is very crucial to note that in expanding the action functionals in \eq{eq} around $\fc$, the same zeroth order term $S[\fc]$ cancels each other in the exponential. As a result, for the in-in path integral there is no need to assume $S[\fc]$ to be finite to have a well defined expansion around $\fc$. This last property, which is also mentioned in the introduction, allows  a more general set of field configurations to become stationary phases.

As one would expect, in the following we will assume $\fc$ to depend only on time,
\be
\fc=\fc(t'),
\ee
which is suitable for cosmological applications. To warm up for our actual construction, we start studying simple models in flat space.\footnote{In flat space the second line of \eq{ara} does not vanish identically. One should impose an extra "Dirichlet" boundary condition at infinity to set $\d\phi^\pm(-\infty)=0$.}

{\it Free massless scalar in flat space:} The field equation $\ddot{\phi}_{cl}=0$ can be solved as $\fc=c_1+c_2t'$.  The only solution obeying \eq{sbc} is $\fc=0$ and thus there is no  solution. 

{\it Free massive scalar in flat space:} The equation of motion $\ddot{\phi}_{cl}+m^2\fc=0$ has oscillating solutions. To impose \eq{sbc} we first keep $t_0$ finite and thus the solution becomes $\fc=c\sin(m(t'-t_0))$. However, $t_0\to-\infty$ limit is not well defined and thus there is no  solution for this case either. 

{\it Massless $\l\phi^4$ theory in flat space:} The equation $\ddot{\phi}_{cl}+\l\fc^3=0$ has  two oscillating solutions which can be expressed in terms of elliptic functions. The solution obeying the necessary boundary condition at $t_0$ (once more keeping $t_0$ finite first) is 
\be\label{fs1}
\fc=\pm c \left(\fr{2}{\l}\right)^{1/4} sn\left[c\left(\fr{\l}{2}\right)^{1/4}(t'-t_0),-1\right],
\ee
where $sn$ is the Jacobi elliptic function, but  $t_0\to-\infty$ limit is ill defined since the argument of $sn$ does not converge. As another try, one may scale the constant $c$ to rewrite the solution as 
\be\label{fs2}
\fc=\pm \fr{c}{t_0} \left(\fr{2}{\l}\right)^{1/4} sn\left[\fr{c}{t_0}\left(\fr{\l}{2}\right)^{1/4}(t'-t_0),-1\right].
\ee
This time, the argument of the $sn$ function has a well defined limit as $t_0\to-\infty$, but one ends up with the trivial solution $\fc=0$ due to the extra factor of $t_0$ appeared in front. 

Although it is possible to consider more examples, these are enough for one to convince himself that in flat space no solution can be found for scalar fields. The reason is that the classical solutions are necessarily oscillating for all times and  \eq{sbc} only fixes the phase of the oscillation. Therefore,  the solution automatically becomes ill defined at time $t$ due to the infinite amount of oscillations it had performed on the way. It is clear that the expansion of the universe, which shows up as a cosmic friction term in the scalar field equation, might change this situation. Let us therefore consider some examples in de Sitter space.

{\it Free massless scalar in de Sitter space:} The equation $\ddot{\phi}_{cl}+3H\dot{\phi}_{cl}=0$ can be solved as $\fc=c_1+c_2\exp(-3Ht')$. To impose \eq{sbc} let us  first keep $t_0$ finite again, which gives $\fc=\phi_0(1-\exp[-3H(t'-t_0)])$. Sending $t_0\to-\infty$, one finds $\fc\to\phi_0$ for all finite $t'$, which is a trivial solution.  

{\it Free massive scalar in de Sitter space:} The equation of motion $\ddot{\phi}_{cl}+3H\dot{\phi}_{cl}+m^2\fc=0$ can be solved by assuming $\fc=\exp(\b t')$ where $\b^2+3H\b+m^2=0$. The solution obeying \eq{sbc} can be written as $\fc=c(\exp[\b_+(t'-t_0)]-\exp[\b_-(t'-t_0)])$. The real part of $\b$ is always negative and therefore $\fc\to0$ as $t_0\to-\infty$ for all finite $t'$, which shows that there is no suitable solution.

{\it Massless $\l\phi^4$ theory in de Sitter space:} In this case, the equation for $\fc$ becomes $\ddot{\phi}_{cl}+3H\dot{\phi}_{cl}+\l\fc^3=0$, which cannot be solved analytically. However, the dynamics is well understood: the Hubble term damps the oscillations about the equilibrium point $\phi=0$ of  $\l\phi^4$ potential. Assuming  initially that the expansion rate is small compared to the frequency of the oscillations, one can write $\phi=Af$, where $A$ represents the slowly varying amplitude and $f$ is the rapidly oscillating function. Taking $A\gg H$, $\dot{A}\sim AH$ and $\dot{f}\gg Hf$ correspond to an expansion that is slow compared to the oscillations. Under these assumptions, the field  equation can be approximately  solved by imposing 
\be
2\dot{A}+3HA=0,\hs{8}\ddot{f}+\l A^2f^3=0,
\ee
which shows that the amplitude decreases exponentially $A=A_0\exp(-3H(t-t_0)/2$ and $f$ is given by the Jacobi elliptic function $sn$. Therefore, the solution can be written as\footnote{The period of the $sn[x,-1]$  is given by the complete elliptic integral of first kind $K(x)$ as $2K(-1)\simeq 2.62$, which shows that the amplitude $A$ in \eq{ds} also fixes the frequency of the oscillations.} 
\be\label{ds}
\fc\simeq  A\left(\fr{2}{\l}\right)^{1/4}sn\left[\left(\fr{\l}{2}\right)^{1/4}\int^t A(t')dt' ,-1\right],
\ee
and it exponentially collapses to zero as  $t_0\to-\infty$. It is easy to see that if the expansion rate is not small compared to the initial oscillation frequency, the solution vanishes more rapidly in the limit. 

One can have a physical understanding of these negative results as follows. As noted before, the in-in path integral is given by the sum over all paths extending from $t_0=-\infty$ to $t$ and then back to $t_0$ again. The trivial path $\phi=0$ corresponding to the "ground state" is clearly a stationary phase of the integral and the perturbation theory works around this path. Consider now the paths with $\phi(t_0)=0$ having arbitrary initial velocities at $t_0$ (such paths exist in the function space provided they satisfy the boundary conditions we discussed in the previous subsection). These paths can be viewed as spontaneous quantum fluctuations around the vacuum and among them only stationary phases have a chance to give significant contribution to the in-in path integral. In flat space, all classical solutions with $\phi(t_0)=0$, i.e. all vacuum fluctuations,  necessarily oscillate indefinitely and their contribution to the path integral averages out zero, which explains the absence of solutions in that case. On the other hand, in de Sitter space fluctuations starting with finite initial velocities at $t_0=-\infty$ are all damped by the expansion of the universe near  infinity so that they all vanish  "before" escaping from the asymptotic region. Therefore, the problem is that the solutions either oscillates too much (flat space) or damped too much (de Sitter space).\footnote{As a side comment let us mention that we are focusing on solutions which has finite velocities $\del \fc(t_0)/\del t'$ so that the second line of \eq{ara} is satisfied. It would be interesting to study solutions with infinite initial velocities still satisfying \eq{ara}, which might beat the infinite damping of the cosmic expansion.}

\begin{figure}
\centerline{\includegraphics[width=7cm]{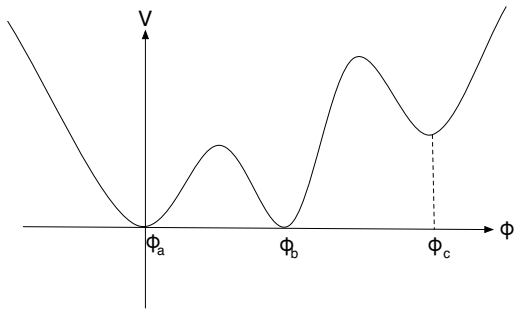}}
\caption{An example of a scalar potential supporting instanton-like solutions.} 
\label{fig2}
\end{figure}

The above comments suggest where to look for nontrivial stationary phases solutions. Take a scalar field propagating in an {\it expanding} FRW space with a potential pictured in Fig. \ref{fig2}. A stationary phase $\fc$ must obey
\be
\ddot{\phi}_{cl}+3\fr{\dot{a}}{a}\dot{\phi}_{cl}+\fr{\del V}{\del \phi}=0.
\ee
Assume that we are expanding around the vacuum $\phi_a=0$. Solutions with small initial velocities are damped around $\phi_a=0$ and the corresponding  solution in the limit $t_0\to-\infty$ is the trivial one $\fc=\phi_a=0$. Consider now the paths with larger initial velocities $\dot{\phi}_{cl}(t_0)>0$ such that the scalar jumps over the first bump in the right and performs damped oscillations around the second vacuum $\phi_b$. It is clear that in the $t_0\to-\infty$ limit the solution becomes\footnote{Note that $t_0\to-\infty$ limit is equivalent to $t'\to+\infty$ limit.}
\be
\fc=\phi_b.
\ee
Similarly, if $\dot{\phi}_{cl}(t_0)$ is chosen large enough so that the scalar overshoots the vacuum $\phi_b$ and starts performing damped oscillations around the locally stable vacuum $\phi_c$, then the limiting solution corresponding to all such initial data becomes 
\be
\fc=\phi_c.
\ee
Thus, all solutions starting from $\fc(t_0)=0$ with finite initial velocities asymptotically becomes either $\fc=\phi_a=0$, $\fc=\phi_b$ or $\fc=\phi_c$, which are the nontrivial  stationary phases. Note that these solutions also satisfy the strong boundary conditions \eq{s}. 

The existence of these  solutions implies that even one starts from the vacuum around $\phi_a=0$, i.e. even when the vacuum wave-functionals in \eq{gb} are chosen accordingly, the generating functional \eq{gb} will get contributions from the vacua $\phi_b$ and $\phi_c$ via the above instanton-like solutions and one should write
\be
Z=Z_a+Z_b+Z_c,
\ee
where $Z_a$, $Z_b$ and $Z_c$ can be calculated using  perturbation theory around the corresponding vacuum. Moreover one finds 
\be
<\phi>=\phi_a+\phi_b+\phi_c,
\ee
thus these solutions change the vacuum expectation value of the scalar from its naive perturbative value. This is very similar to tunneling to the perturbatively inaccessible vacua. 

\subsection{The case of finite $t_0$}

For finite $t_0$, there appears additional technical difficulties. One of the main problems is that in the absence of an asymptotic region even the free vacuum cannot  be uniquely determined in an expanding universe \cite{un} (recall how one defines the Bunch-Davies vacuum). On the other hand, in an interacting theory one cannot use the trick of giving a small imaginary piece to time coordinate to project onto the exact vacuum and thus even perturbation theory might become difficult to apply. Moreover, as we will see in a moment, the existence of stationary phases  depends on whether one imposes the strong or the weak boundary conditions,  discussed above. 

We can bypass some of these difficulties since we are employing a semiclassical approximation. Taking  $\phi=0$ as the minimum of the scalar potential, it is reasonable to assume that the vacuum wave-functionals are oscillating functionals of the field variables around this minimum. Therefore, in the stationary phase approximation, the presence of the vacuum wave-functionals in the path integral implies $\fc^\pm(t_0)=0$ as a result of $\phi^\pm(t_0)$ integrals in \eq{i}. The variation of the phase in the path integral is still given by \eq{ara} (with finite $t_0$) and setting the coefficient of the first line to zero again implies $\fc^+=\fc^-$. Therefore, stationary phases must  obey
\be
\fc^+=\fc^-\equiv\fc,\hs{8}\fc(t_0)=0,
\ee
similar to $t_0=-\infty$ case. 

The rest of the discussion depends on which boundary conditions are imposed in the function space. If one assumes the weak boundary conditions \eq{w}, then  the only way to set the second line of \eq{ara} to zero is to impose  $\del\fc(t_0)/\del t'=0$ (note that $\d\phi^\pm(t_0)$ variations are independent), which then gives $\fc=0$. Thus, there is no  solution for weak boundary conditions.

On the other hand, to satisfy the second strong condition involving $a(t_0)$ in \eq{s} one should  impose Dirichlet or  Neumann conditions for $+$ and $-$ branches. When the Neumann condition is chosen either for $\phi^+$ or $\phi^-$, then one gets $\fc=0$. Therefore, instanton-like solutions  exist only when Dirichlet conditions are imposed, i.e. $\phi^\pm(t_0)=0$. In that case the second line of \eq{ara} is also satisfied since $\d\phi^\pm(t_0)=0$ and as a result one finds
\be \label{ft0}
\fc(t_0)=0,\hs{8}\fr{\del\fc(t)}{\del t'}=0,
\ee
where the second condition follows from the first set of strong boundary conditions in \eq{s}. 

It is not difficult to construct  nontrivial classical solutions\footnote{Note that the function space with strong boundary conditions is smaller than the function space with weak boundary conditions. Therefore, one should not surprise to see that the same path integral has stationary phases in the first space but not in the second one.}  (actually infinitely many of them) satisfying \eq{ft0}. As an example consider again massless $\l \phi^4$ theory in flat space. The solution \eq{fs1} already satisfies the condition $\fc(t_0)=0$ and imposing the second one in \eq{ft0} gives
\be\label{fsi}
\fc^{(n)}(t')=\pm \left(\fr{2}{\l}\right)^{1/2}\fr{(2n+1)K(-1)}{t-t_0} sn\left[\fr{(2n+1)K(-1)}{t-t_0}(t'-t_0),-1\right],
\ee
where $K(x)$ is the elliptic integral of first kind, $K(-1)$ equals the half  period of the Jacobi elliptic function $sn[x,-1]$ and $n$ is an integer. Note that as  $n$ getting larger, both the amplitude and the frequency of the oscillations grow, which shows the necessity of imposing an upper limit (cutoff) for $n$.  The vacuum expectation value of the scalar still vanishes  due to $\phi\to-\phi$ symmetry (note $\pm$ signs in \eq{fsi}):
\be
<\phi>=0. 
\ee
Note also that as $t_0\to-\infty$, $\fc\to 0$ consistent with our earlier considerations. 

Expanding the theory around one of these solutions gives the potential
\be\label{tp}
\hat{V}=3\l\, \fc^{(n)}(t')^2\,\hat{\phi}^2+2\l\fc^{(n)}(t')\,\hat{\phi}^3+\fr{\l}{2}\hat{\phi}^4,
\ee
where $\hat{\phi}$ is the fluctuation field. The potential contains time dependent coupling constants and specifically a time dependent mass term. Since $\fc^{(n)}$ oscillates about zero, the shape of the potential also changes in time due to sign flips of the cubic term. From \eq{fsi}, the mass term becomes independent of $\l$ and the cubic interaction term has the strength $\sqrt{\l}$, which cannot arise in any perturbative expansion in $\l$. 

Similar solutions can be seen to exist for different scalar potentials and for scalars propagating in an expanding FRW universe, although it might not be possible to write down analytical expressions. For instance, in the massless $\l\phi^4$ theory defined in an exponentially expanding spacetime (which is not  de Sitter space since $t_0$ is finite), the solution become very much like \eq{ds}, where the initial value of the  time dependent amplitude must be quantized to satisfy \eq{ft0}. Indeed,  \eq{ds} becomes more and more reliable for larger amplitudes and the solution approaches to \eq{fsi}. 

\section{Conclusions}

Path integral formulation gives valuable non-perturbative information about the structure of quantum  field theories, which is otherwise hard, if not impossible, to acquire. Likewise, it is not going to be surprising to see that in-in path integrals will reveal some non-perturbative aspects of quantum contributions to cosmological correlations. Despite its evident importance, the path integral formalism applied to  in-in correlation functions is not studied too much. Although one may encounter earlier works like \cite{e00,e01}, (to our knowledge) even the very basic result on the equivalence of the operator and the path integral approaches in perturbation theory has been proved relatively recently  in \cite{w1}. Similarly, a quantitative in-in path integral treatment of the cosmological perturbation theory has been given about two years ago \cite{pr}, which is very novel. Our work confirms that in-in path integral formalism will be an important technique in searching for non-perturbative effects for cosmological correlation functions and some surprising results may still emerge. 

Including the contributions of  non-trivial stationary phases to in-in path integrals, some perturbative quantum results are destined  to change and this might have important theoretical and observational consequences. For example, in discussing the effects of symmetry breaking and  Higgs mechanism in cosmology,  the existence of instanton-like states must definitely be considered since they allow all locally stable minima to contribute the cosmological correlations. Namely, even a gauge symmetry is broken by a non-zero vacuum expectation of a scalar in a certain cosmological era, correlation functions might still be modified by the presence of the symmetric vacuum via instanton-like states.  On the other hand, if time dependent  solutions (i.e. instanton-like solutions whose time dependence survives the cosmic damping) exists similar to \eq{fsi}, they might affect the spectrum of cosmological perturbations (like by changing the tilt of the spectrum)  since the spectrum is sensitive to shape of the scalar potential and these solutions give rise to potentials with time dependent coupling constants like  \eq{tp}.

It is possible to extend this work in  different directions. It would be very interesting to apply the present arguments  to a quantum mechanical problem to test the validity of the stationary phase approximation in calculating in-in path integrals. As an inevitable extension, one can consider gauge fields coupled to scalars and work out  how the existence of instanton-like solutions might have an impact on the Higgs mechanism on cosmological scales. It would also be an important exercise to include the gravitational degrees of freedom and determine the semiclassical gravitational contributions, which might affect the standard cosmological perturbation theory. Of course the problems involving the gauge and the gravitational fields  have an extra complication related to gauge degrees of freedom, but it is worth to understand how the standard picture will be modified. In that way one might be more confident about inflation or find a caveat  in its main arguments. 

\begin{acknowledgments}

I would like to thank Emre Onur Kahya and Vak\i f Kemal \"{O}nemli for useful discussions. This work is supported by T\"{U}B\.{I}TAK B\.{I}DEB-2219 grant. 

\end{acknowledgments}

\end{document}